\documentclass[10pt, conference, compsocconf]{IEEEtran}

\usepackage{paralist}
\usepackage{amsmath}
\usepackage{url}
\usepackage{multirow}

\usepackage{tikz}
\usetikzlibrary{chains}

\usepackage{listings}
\usepackage{color}

\usepackage{verbatim}

 \usepackage{pgf-umlsd}

\usepackage{xwatermark}
\usepackage{pgfplots}

\definecolor{codegreen}{rgb}{0,0.6,0}
\definecolor{codegray}{rgb}{0.5,0.5,0.5}
\definecolor{codepurple}{rgb}{0.58,0,0.82}
\definecolor{backcolour}{rgb}{0.95,0.95,0.92}
\definecolor{darkblue}{rgb}{0.0,0.0,0.6}

\definecolor{mGreen}{rgb}{0,0.6,0}
\definecolor{mGray}{rgb}{0.5,0.5,0.5}
\definecolor{mPurple}{rgb}{0.58,0,0.82}
\definecolor{backgroundColour}{rgb}{0.95,0.95,0.92}
\definecolor{backgroundColour}{rgb}{0.95,0.95,0.92}

\lstdefinestyle{mystyle}{
  commentstyle=\foonotesize\color{codegreen},
  backgroundcolor=\color{backcolour},
  stringstyle=\color{codepurple},
  basicstyle=\footnotesize,
  breakatwhitespace=false,
  breaklines=true,
  captionpos=b,
  keepspaces=true,
  showspaces=false,
  showstringspaces=false,
  showtabs=false,
  tabsize=2,
  numberstyle=\tiny\color{mGray},
  numbers=left,                    
  numbersep=5pt,    
  float=tp,
  floatplacement=tbp
}

\newif\ifComments
\Commentstrue
\newcommand{\REM}[1]{}

\ifComments
\newcommand{\herve}[1]{\noindent\textcolor{pink}{Herve: {#1}}}
\newcommand{\juan}[1]{\noindent\textcolor{violet}{juan: {#1}}}
\newcommand{\marcio}[1]{\textcolor{green}{ {#1}}}
\newcommand{\ramon}[1]{\noindent\textcolor{orange}{Ramon: {#1}}}
\newcommand{\guido}[1]{\noindent\textcolor{magenta}{Guido: {#1}}}
\newcommand{\rem}[1]{\noindent\textcolor{brown}{Removed: {#1}}}

\newcommand{\ed}[1]{\noindent\textcolor{red}{ {#1}}}
\else
\newcommand{\herve}[1]{}
\newcommand{\juan}[1]{}
\newcommand{\marcio}[1]{}
\newcommand{\ramon}[1]{}
\newcommand{\guido}[1]{}
\newcommand{\rem}[1]{}

\newcommand{\ed}[1]{}
\fi 

\newcommand{\rsec}[1]{Section~\ref{sec:#1}}

\newcommand{\rfig}[1]{Figure~\ref{fig:#1}}

\newcommand{\rlst}[1]{Listing~\ref{lst:#1}}

\newcommand{\tit}[1]{{\textit{#1}}}

\usepackage{booktabs}   %% For formal tables:
\usepackage{subcaption} %% For complex figures with subfigures/subcaptions

\begin{document}
%
% paper title
% Titles are generally capitalized except for words such as a, an, and, as,
% at, but, by, for, in, nor, of, on, or, the, to and up, which are usually
% not capitalized unless they are the first or last word of the title.
% Linebreaks \\ can be used within to get better formatting as desired.
% Do not put math or special symbols in the title.
\title{Enabling OpenMP Task Parallelism on Multi-FPGAs}

% author names and affiliations
% use a multiple column layout for up to three different
% affiliations
\begin{comment}
\author{
  \IEEEauthorblockN{
    Anonymous authors. 
  }
  \IEEEauthorblockA{
    \ \\
    \ \\
    \ 
  }
}
\end{comment}

%\begin{comment}
\author{
  \IEEEauthorblockN{
    Ramon Nepomuceno,
    Renan Sterle,
    Guilherme Valarini,
    Marcio Pereira, 
    Hervé Yviquel, and
    Guido Araujo
  }
  \IEEEauthorblockA{
    Institute of Computing, University of Campinas, Brazil\\
    \{ramon.nepomuceno, mpereira, herve, guido\}@ic.unicamp.br\\
    \{r176536, g168891\}@dac.unicamp.br
  }
}
%\end{comment}

% make the title area
\maketitle

% As a general rule, do not put math, special symbols or citations
% in the abstract
\begin{abstract}
FPGA-based hardware accelerators have received increasing attention mainly due to their ability to accelerate deep pipelined applications, thus resulting in higher computational performance and energy efficiency. Nevertheless, the amount of resources available on even the most powerful FPGA is still not enough to speed up very large modern workloads. To achieve that, FPGAs need to be interconnected in a Multi-FPGA architecture capable of accelerating a single application. However, programming such architecture is a challenging endeavor that still requires additional research. This paper extends the OpenMP task-based computation offloading model to enable a number of FPGAs to work together as a single Multi-FPGA architecture. Experimental results for a set of OpenMP stencil applications running on a Multi-FPGA platform consisting of 6 Xilinx VC709  boards interconnected through fiber-optic links have shown close to linear speedups as the number of FPGAs and IP-cores per FPGA increase. 
\end{abstract}

\IEEEpeerreviewmaketitle

\section{Introduction}
With the limits imposed by the power density of semiconductor technology, heterogeneous systems became a design alternative that combines CPUs with domain-specific accelerators to improve power-performance efficiency \cite{oneal:surveypower:2018}. A modern heterogeneous system typically combines general-purpose CPUs and GPUs to speedup  complex scientific applications\cite{guo:gpu:2020}. However, for many specialized applications that can benefit from pipelined parallelism (e.g. FFT, Networking), FPGA-based hardware accelerators have shown to produce improved power-performance numbers  \cite{ghazawi:promise:2008, lee:openaccfpga:2016, strickland:fpgahpc:2018, reichenbach:hetfpga:2019}. Moreover, FPGA's reconfigurability facilitates the adaptation of the accelerator to distinct types of workloads and applications. In order to leverage on this, cloud service companies like Microsoft Azure~\cite{caulfield:azure:2016} and Amazon AWS \cite{awsf1} are offering heterogeneous computing nodes with integrated FPGAs.

Given its small external memory bandwidth \cite{cong:fxg:2018} FPGAs do not perform well for applications that require intense memory accesses. Pipelined FPGA accelerators \cite{jiang:dnnfpgas:2019, farooq:routing:2018, azeem:fpgasdebug:2016} have been designed to address this problem but such designs are constrained to the boundaries of a single FPGA or are limited by the number of FPGAs that it can handle. By connecting multiple FPGAs, one can design deep pipelines that go beyond the border of one FPGA, thus allowing data to be transferred through optical-links from one FPGA to another without using external memory as temporal storage. Such deep pipeline accelerators can considerably expand the application of FPGAs, thus enabling increasing speedups as more FPGAs are added to the system \cite{waidyasooriya:fpgastencil:2019}.

Unfortunately, programming such Multi-FPGA architecture is a challenging endeavor that still requires additional research~\cite{kunzman:proghet:2011, pu:prgdsl:2017}. Synchronizing the accelerators inside the FPGA and seamlessly managing data transfers between them are still significant design challenges that restrict the adoption of such architectures. This paper addresses this problem by extending the LLVM OpenMP task programming model~\cite{openmp45} to Multi-FPGA architectures.

\begin{figure}[!t]
    \centering
    \includegraphics[width=\columnwidth]{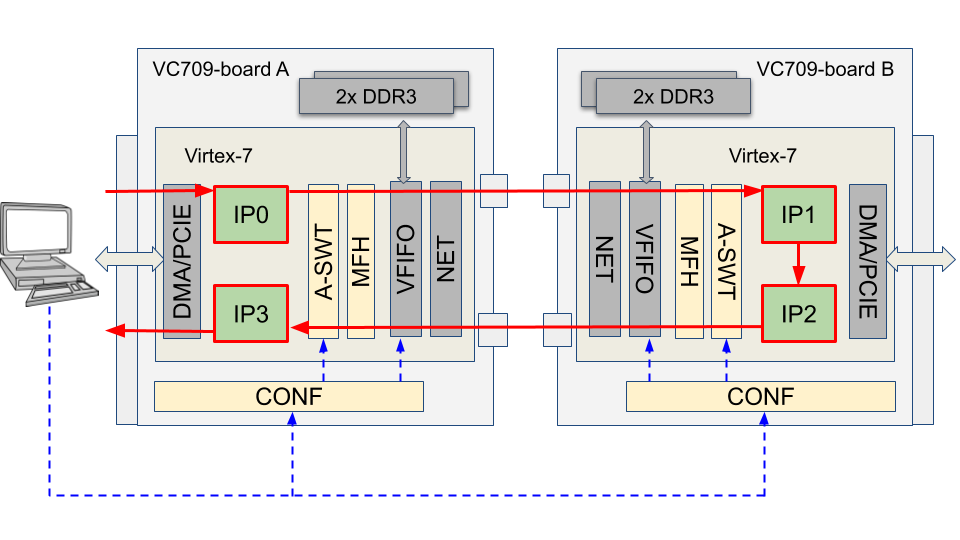}
    \caption{An optical-link interconnected Multi-FPGA architecture running an OpenMP pipelined application.}
    \label{fig:cluster_inc}
\end{figure}

At a higher abstraction level, the programming model proposed in this paper enables the programmer to see the FPGA as a regular OpenMP device and the IP-cores accelerators (IPs) as OpenMP tasks. In the proposed approach, the OpenMP task dependence mechanism transparently coordinates the IPs' work to run the application. For example,  consider the simple problem of processing the elements of a vector V in a pipeline fashion using four IPs (IPO-IP3) programmed in two FPGAs. Each IPi (i = 0-3) performs some specific computation \textit{foo(V,i)}. The Multi-FPGA architecture used in this example is shown \rfig{cluster_inc} and contains two VC709\footnote{A similar approach can also be used for modern Alveo FPGAs.}  FPGA boards interconnected by two fiber-optics links. Each FPGA is programmed with: (a) a module for communication with the PCIE interface (DMA/PCIE); (b) two IPs from the set IP0-IP3; (c) a NET module for communication with the optical fibers; (d) a Virtual FIFO module (VFIFO) for communication with memory; (e) a MAC frame handler (MFH) to pack/unpack data; and (f)  a packet switch (A-SWT) module capable of moving data among the IPs, even if they seat on FPGAs from two distinct boards. As shown in the figure, the vector is initially moved from the host memory (left of \rfig{cluster_inc}) and then pushed through the IP0-IP3 pipeline, returning the final result into the host memory.

The OpenMP program required to execute this simple application uses just a few code lines (see \rlst{inc_code}). Its simplicity is possible due to the techniques proposed in this paper, which leverage on the OpenMP task dependence and computation offloading abstraction to hide the complexity needed to move the vector across the four IPs. The proposed techniques extend the OpenMP runtime with an FPGA device plugin which: (a) maps task data to/from the FPGAs; (b) transparently handles the data dependencies between IPs located in distinct FPGAs; and (c) eases the synchronization of IPs' execution.

The main contributions of this paper are summarized below:
\begin{itemize}
   \item A new Clang/LLVM plugin that understands FPGA boards as OpenMP  devices, and uses OpenMP \textit{declare variant} directive to specify hardware IPs;
   \item A mechanism based on the OpenMP task dependence and computation offloading model that enables transparent communication of IPs  in a Multi-FPGA architecture;
   \item A programming model based on OpenMP task parallelism, which makes it simple to move data among FPGAs, CPUs or other acceleration devices (e.g. GPUs), and that allows the programmer to use a single programming model to run its application on a truly heterogeneous architecture. 
  \end{itemize}

The rest of the paper is organized as follows. \rsec{background} discusses background material and  \rsec{mult_fpgas} details the software and hardware mechanisms that were designed to enable OpenMP task parallelism on a Multi-FPGA architecture. \rsec{stencil} shows how this architecture can be used to design a scalable stencil pipeline application. \rsec{experiments} describes the experimental setup and analyzes their results. \rsec{relwork} discusses related works and finally \rsec{conclusions} concludes the work. 

\section{Background} 
\label{sec:background}
This section provides background material required for understanding the proposed approach. \rsec{omp} reviews the basic concepts of the OpenMP accelerator programming model, while \rsec{thefpga} details the Xilinx modules used to assemble the testing platform.

\subsection{OpenMP Accelerator Model}
\label{sec:omp}

The OpenMP specification \cite{openmp2008book} defines a programming model based on code annotations that use minimal modifications of the program to expose potential parallelism. OpenMP has been successfully used to parallelize programs on CPUs and GPUs residing on a single computer node.

Tasks were introduced to OpenMP in version 3.0 \cite{openmp30}, to expose a higher degree of parallelism through the non-blocking execution of code fragments annotated with \textit{task} directives. The annotated code is outlined as tasks that are dispatched to the OpenMP runtime by the \textit{control thread}. Dependencies between tasks are specified by the \textit{depend} clause, which the programmer uses to define the set of input variables that the task depends on and the output variables that the task writes to.  Whenever the dependencies of a given task are satisfied, the OpenMP runtime dispatches it to a \textit{ready queue} which feeds a pool of \textit{worker threads}. The OpenMP runtime manages the task graph, handles data management, creates and synchronizes threads, among other activities.

\begin{lstlisting}[style=mystyle, label={lst:tasks}, caption={OpenMP tasks in CPUs.}]
int V[M];
bool deps[N+1];
#pragma omp parallel
#pragma omp single
for(int i=0; i<N; i++){
  #pragma omp task \
  depend(in:deps[i]) depend(out:deps[i+1])
  foo(V,i);
}
\end{lstlisting}

\begin{lstlisting}[style=mystyle, label={lst:target_depend}, caption={Offloading OpenMP tasks to GPU.}]
int V[M];
bool deps[N+1];
#pragma omp parallel
#pragma omp single
for(int i=0; i < N; i++){
  #pragma omp target device(GPU) \ 
  depend(in:deps[i]) depend(out:deps[i+1]) \
  map(tofrom:V[:M]) nowait
  foo(V,i);
}
\end{lstlisting}

Listing \ref{lst:tasks} shows an example of a simple program that uses OpenMP tasks to perform a sequence of increments in the vector V, similarly as discussed in \rfig{cluster_inc}. First, the program creates a pool of \textit{worker threads}. This is done using the \textit{\#pragma omp parallel} directive (line 3). It is with this set of threads that an OpenMP program runs in parallel. Next, the \textit{\#pragma omp single} directive (line 4) is used to select one of the \textit{worker threads}. This selected thread is called \textit{control thread} and is responsible for creating the tasks.  The \textit{\#pragma omp task} directive (line 6) is used to effectively build N \textit{worker tasks}, each one  encapsulating a call to function \textit{foo(V,i)} (line 8) that performs some specific operation on the vector V depending on the task (loop) index. Worker threads are created  by the OpenMP runtime to run on the cores of the system. In order to assure the execution order of the tasks the \textit{depend} clause (line 7) is used to specify the input and output dependencies of each task. In the specific case of the example of Listing \ref{lst:tasks} we have to introduce a \textit{deps} vector which assures that the \textit{foo(V,i)} tasks are executed in a pipeline order.

Now assume the user wants to accelerate the above described tasks in a GPU. For this purpose, consider the code in \rlst{target_depend} which is the version of \rlst{tasks} using a GPU as an acceleration device. In  \rlst{target_depend} the \textit{target} directive (line 6) has a similar role as the \textit{task} directive  from \rlst{tasks}, but has for goal to offload the computation to an acceleration device instead of a CPU. The \textit{device} clause (line 6) receives as an argument an integer which specifies the device that will perform the computation. In the case of \rlst{target_depend} this number corresponds to a GPU. Similar to the \textit{task} directive, the \textit{target} directive also accepts the \textit{depend} clause (lines 7). The \textit{map} clause (line 8) specifies the direction of the transfer operation that the host must execute with the given data. The most common transfer operations are \textit{to}, \textit{from}, and \textit{tofrom}. For example, in line 8 of \rlst{target_depend} vector V is annotated with \textit{tofrom} given that it is first sent from the host memory to the GPU memory, and then  received back after processed by function \textit{foo}. Notice that clause \textit{nowait} also appears in line 8 of Listing \ref{lst:tasks}. This clause is necessary because, by default, the \textit{target} directive is blocking. In other words, the \textit{nowait} clause allows the control thread to create all tasks without waiting for the previous ones to complete. Finally in line 9 of \rlst{target_depend} the worker thread \textit{foo} function is called.

The current LLVM OpenMP implementation of task offloading is not enough to handle devices like Multi-FPGA. This paper proposes an OpenMP plugin implementation for Multi-FPGA devices. Moreover, it extends the LLVM OpenMP runtime to substitute software tasks for \textit{target} tasks running on hardware IPs of a Multi-FPGA architecture. This is done by implementing the \textit{declare variant} pragma which is already defined in the OpenMP standard. Also, the \textit{depend} and the \textit{map} clauses are used to dynamically create communication paths between the IPs of the various FPGAs in the architecture.

\subsection{The Hardware Platform}
\label{sec:thefpga}

\begin{figure}[t!]
    \centering
    \includegraphics[width=\columnwidth]{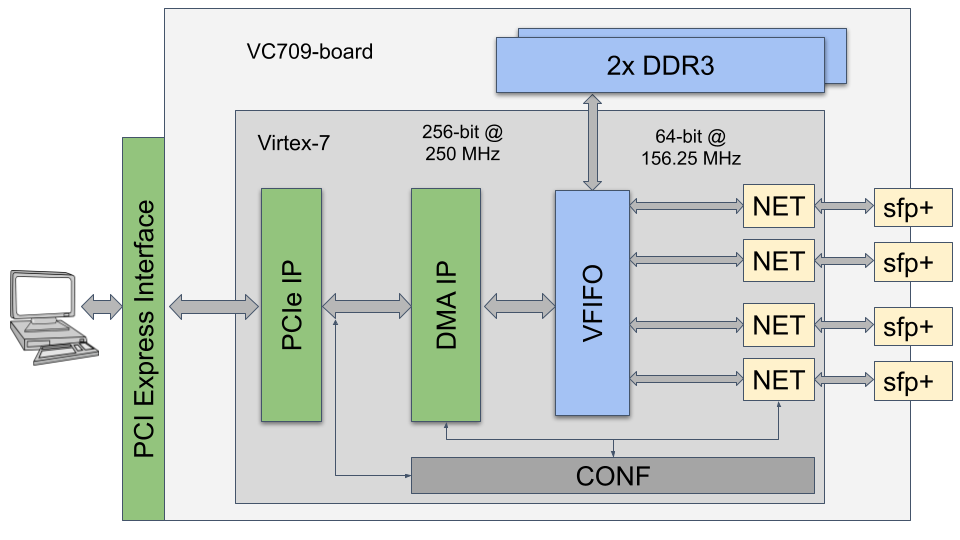}
    \caption{VC709 Target Reference Design.}
    \label{fig:vc709}
\end{figure}

The Xilinx VC709 Connectivity Kit provides a hardware environment for developing and evaluating designs targeting the Virtex-7 FPGA. The VC709 board provides features common to many embedded processing systems, including dual DDR3 memories, an 8-lane PCI Express interface \cite{budruk:PES:2003}, and small SFP connectors for fiber optical (or Ethernet) transceivers. In addition to the board's physical components, the kit also comes with a Target Reference Design (TRD) featuring a PCI Express IP, a DMA IP, a Network Module, and a Virtual FIFO memory controller interfacing to DDR3 memory. Figure \ref{fig:vc709} shows the schematic of the main board components and their corresponding components.

This kit was selected to explore the ideas discussed in this paper due to its reduced cost and the fact that the TRD has components ready for inter-FPGAs communication. Each one of the TRD components is explained in more detail below.

\noindent{\bf{PCIE and DMA}.} The PCI Express IP (Figure \ref{fig:vc709}) provides a wrapper around the integrated block in the FPGA. The wrapper combines the Virtex-7 XT FPGA Integrated Block for PCIe with transceivers, clocking, and reset logic. It also provides an AXI4-Stream interface to the host.

\noindent{\bf CONF.}  The Configuration Registers (Figure \ref{fig:vc709}) are used to read and write control/status information to/from the FPGA components. This information ranges from configuring the network module to reading performance, power, and temperature information. There is also a free address range available that is used to store configuration information to the specific IPs used in this work.

\noindent{\bf VFIFO.} The TRD uses DDR3 space to implement a Virtual FIFO (VFIFO Figure \ref{fig:vc709}). This VFIFO is used to avoid back-pressure to the PCIe/DMA modules.

\noindent{\bf Network Subsystem}. This subsystem is composed of four NET modules  (see Figure \ref{fig:vc709}) containing each XGEMAC module and logic required to enable the physical layer of the Multi-FPGA communication network. Each XGEMAC module receives data in the MAC frame format and sends it to the physical layer. Each NET module is connected to an SFP port capable of handling 10Gb/s per channel, resulting in a total of 40Gb/s bandwidth for the board.

\section{The OpenMP Multi-FPGA Architecture} 
\label{sec:mult_fpgas}

The approach proposed in this paper has two main goals: (a) to extend the LLVM OpenMP runtime so it can recognize FPGA IPs as OpenMP tasks (i.e.\textit{targets}) and; (b) to design a hardware mechanism that enables transparent communication of IP data dependencies through the cluster of FPGAs. The following two sections detail how these two goals have been achieved.

\subsection{Extending OpenMP} 

To explain how the proposed system works, please consider the code fragment shown in Listing \ref{lst:inc_code}. The \textit{declare variant} directive (line 1), which is part of the OpenMP standard, declares a specialized hardware (FPGA IP-core) variation (\textit{hw\_laplace2d}, in line 4) of a C function (\textit{do\_laplace2d}, in line 2), and specifies the context in which that variation  should be called. For example, line 3  of Listing \ref{lst:inc_code} states that the variant \textit{hw\_laplace2d} should be selected when the proper \textit{vc709} device flag is provided to the compiler at compile time. This compiler flag is matched by the \textit{match (device=arch(vc709))} clause and when the call to function \textit{do\_laplace2d(\&V, x, y)} (line 16) is to be executed a call to the IP-core \textit{hw\_laplace2d(int*, int, int)} is performed instead. 

As shown in line 11 of Listing \ref{lst:inc_code}, the \textit{main} function creates a pipeline of N tasks. Each task receives a vector V containing \textit{h*w} (height and width) grid elements that are used to calculate a Laplace 2D stencil. As the tasks are created within a target region and the \textit{vc709} device flag was provided to the compiler, the \textit{hw\_laplace2d} variant is selected to run each task in line 16. The compiler then uses this name to specify and offload the hardware IP that will run the task. As a result, at each loop iteration, a hardware  IP task is created inside the FPGA. Details of the implementation of the Laplace 2D IP and other stencil IPs used in this work are provided in Section \ref{sec:stencil}.

The \textit{map} clause (line 12) specifies that the data is mapped back and forth to the host at each iteration. However, the implemented mapping algorithm concludes that vector V is sent to the IP from the host memory and its output forwarded to the next IP in the following iteration. The interconnections between these IPs are defined according to the \tit{depend} clauses (line 13). In the particular example of Listing \ref{lst:inc_code}, given that the dependencies between the tasks follow the loop iteration order, a simple pipelined task graph is generated.

\begin{lstlisting}[style=mystyle, label={lst:inc_code}, caption={Offloading task computations to FPGA IPs.}]
#pragma omp declare variant 
  (void do_laplace2d(int*,int,int))  \
  match (device=arch(vc709))               
extern void hw_laplace2d(int*,int,int);  

int main(){
  float  V[h*w];
  bool deps[N+1];
  #pragma omp parallel
  #pragma omp single
  for (int i = 0; i < N; i++){
    #pragma omp target map(tofrom:V[:(h*w)]) \
    depend(in:deps[i]) depend(out:deps[i+1]) \
    nowait
    {
      do_laplace2d(&V,h,w);
    }
  }
}
\end{lstlisting}

A careful comparison of \rlst{inc_code} with \rlst{target_depend} reveals that in terms of syntax, the proposed approach does not require the user to change anything concerning the OpenMP standard, besides specifying the \textit{vc709} flag to the compiler. This gives the programmer a powerful verification flow. He/she can write the software version of \textit{do\_laplace2d} for algorithm verification purpose, and then switch to the hardware (FPGA) version \textit{hw\_laplace2d} by just using the \textit{vc709} compiler flag. 

To achieve this level of alignment with the OpenMP standard, two extensions were required to the OpenMP runtime implementation: (a) a modification in the task graph construction mechanism and; (b) the design of the VC709 plugin in the libomptarget library. All these changes have been done very carefully to ensure full compatibility with the current implementation of the OpenMP runtime.

\noindent{\bf Managing the Task Graph}. The first modification made to the OpenMP runtime has to do with how it handles the task graph. In the current OpenMP implementation the graph is built and consumed at runtime. Whenever a task has its dependencies satisfied, it is available for a worker thread to execute. After the worker thread finishes, the task output data is sent back to the host memory. This approach satisfies the needs of a single accelerator, but  causes unnecessary  data movements for an Multi-FPGA architecture as the output data of one (FPGA) task IP may be needed as input to another task IP. To deal with this problem, the OpenMP runtime was changed so that tasks are not immediately dispatched for execution as they are detected by the control thread. In the case of FPGA devices, the runtime waits for the construction of the task graph at the synchronization point at the end of the scope of the OpenMP \textit{single} clause (line 18 of \rlst{inc_code}).

\begin{figure}[!t]
    \centering
    \includegraphics[width=\columnwidth]{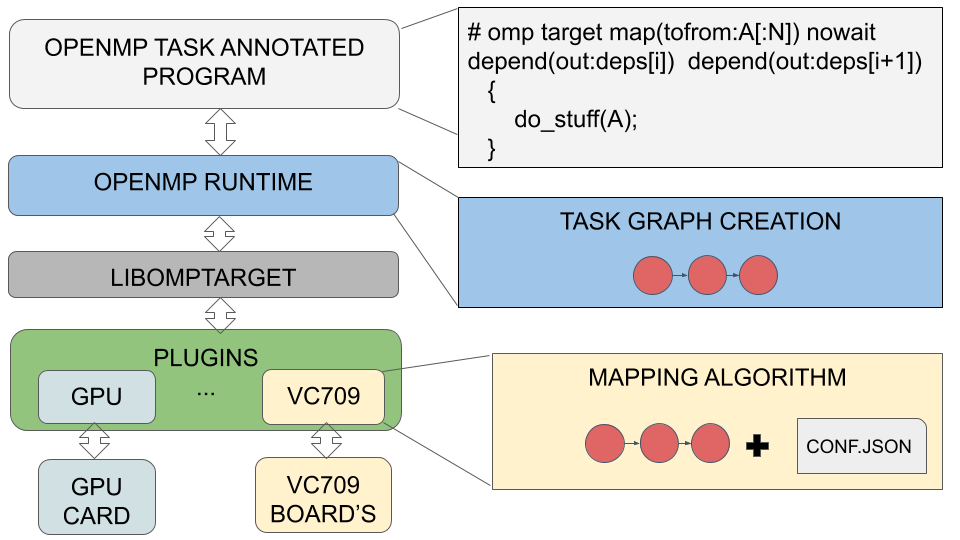}
    \caption{OpenMP software stack with VC709 plugin.}
    \label{vc709_pluging}
\end{figure}

\noindent{\bf Building the VC709 Plugin}. In the OpenMP implementation of the Clang/LLVM compiler \cite{lattner:LCF:2004}, kernel/data offloading are performed by a library called \textit{libomptarget} \cite{bertolli:IGS:2015}. This library provides an agnostic offloading mechanism that allows the insertion of a new device to the list of devices that the OpenMP runtime supports and is responsible for managing kernel and data offloading to acceleration devices. Therefore, to allow the compiler to offload to the VC709 board, it was necessary to create a plugin in this library. Figure \ref{vc709_pluging} illustrates where the plugin is located in the software stack. 

As shown in Figure \ref{vc709_pluging}, the plugin receives the task graph generated by the runtime and maps these tasks to the available IPs in the cluster. The cluster configuration is passed through a \textit{conf.json} file, which contains: (a) the location of the bitstream files, (b) the number of FPGAs, (c) the IPs available in each FPGA, and (d) the addresses of IPs and FPGAs.  As in our experiments, the FPGAs are connected in a ring topology, a round-robin algorithm is used to map tasks to IPs. Each task is mapped in a circular order to the free IP that is closest to the host computer.

\subsection{Designing the Hardware Platform} 
\label{sec:hw}

Besides the extensions to OpenMP, an entire hardware infrastructure was designed to support OpenMP programming of Multi-FPGA architectures. This infrastructure leverages on the Target Reference Design (TRD) presented in Section \ref{sec:thefpga}, but could also be ported to other modern Alveo FPGAs. To facilitate its understanding, the design is described below using two perspectives: Single-FPGA execution and Multi-FPGA execution.

\noindent{\bf Single-FPGA Execution}. As discussed above, when the acceleration device is an FPGA, OpenMP considers the FPGA IPs as OpenMP tasks which are specified by the name of a predefined variant function.

Consider, for example, the hardware infrastructure in Figure \ref{fig:cluster_inc}, but this time with a single FPGA board and 4 task IPs (IP0-IP3).  These IPs are separately designed using a standard FPGA toolchain (e.g. Vivado), which is not addressed in this paper. To connect the IPs into the infrastructure, the designer just needs to ensure that they use the AXI-Stream interface. The infrastructure can be easily changed to accept other interfaces, although for the purpose of this paper AXI-Stream suffices.

According to the proposed programming model, FPGA IPs can execute tasks that have dependencies between each other. In order to implement such dependencies an \textit{AXI4-Stream Interconnet} (AXIS-IC)  module \cite{AXI4Stre90:online} was inserted to the infrastructure (Figure \ref{fig:cluster_inc}). This enables the   IPs to communicate directly to each other, based on the OpenMP dependencies programmed between them, thus avoiding unnecessary communication through the host memory.

The \textit{AXI4-Stream Interconnect} is an IP core that enables the connection of heterogeneous master/slave AMBA® AXI4-Stream protocol compliant endpoint IP. The AXI4-Stream Interconnect routes data from one or more AXI4-Stream master channels to one or more AXI4-Stream slave channels. The VC709 plugin uses the \textit{CONF} register (Figure \ref{fig:cluster_inc}) bank to program the source and destination ports of each IP according to their specified task dependencies.

\noindent{\bf Multi-FPGA Cluster Execution}. A Multi-FPGA architecture is composed of one or more cluster nodes containing at least one FPGA board each. To enable such architecture, routing capability needs to be added to each FPGA so that IPs from two different boards or nodes communicate through the optical links.

To enable that, a MAC Frame Handler module (MFH) was designed and inserted into the hardware infrastructure, as shown in Figure \ref{fig:cluster_inc}. This module is required because the \textit{Network Subsystem} that routes packages through the  optical fibers that connects the boards receives data in the form of MAC Frames, which contain four fields: (a) \textit{destination},  (b) \textit{source}, (c)  \textit{type/length} and (d)  \textit{payload}. Therefore, to use the optical fibers, a  module that can assemble and disassemble MAC frames is required. 

The MFH module is responsible for inserting and removing the source and destination MAC addresses and  \textit{type/lengh} fields whenever the IPs need to send/receive data through the \textit{Network Subsystem}. MAC addresses are extracted from the dependencies in the task graph while the \textit{type/lengh} fields are extracted from the \textit{map} clause. The VC709 plugin uses this information to set up the \textit{CONF} registers, which in turn configure the MFH module.
\begin{table}[!h]
\begin{tabular}{|c|c|}
\hline
Kernel & Computations                                                                         \\ \hline
1           & \begin{tabular}[c]{@{}c@{}}$ 0.25(V^t_{i,j-1}+V^t_{i-1,j}+V^t_{i+1,j}+V^t_{i,j+1})  $\end{tabular} \\ \hline
2           & \begin{tabular}[c]{@{}c@{}}$C_1.V^t_{i,j-1}+C_2.V^t_{i-1,j}+C_3.V^t_{i,j}+C_4.V^t_{i+1,j}+C_5.V^t_{i,j+1}$\end{tabular} \\ \hline
3           & \begin{tabular}[c]{@{}c@{}}$C_1.V^t_{i-1,j-1}+C_2.V^t_{i,j-1}+C_3.V^t_{i+1,j-1}+$\\$C_4.V^t_{i-1,j}+C_5.V^t_{i,j}+C_6.V^t_{i+1,j}+$\\ $C_7.V^t_{i-1,j+1}+C_8.V^t_{i,j+1}+C_9.V^t_{i+1,j+1}$\end{tabular} \\ \hline
4           & \begin{tabular}[c]{@{}c@{}}$0.25(V^t_{i,j-1,k}+V^t_{i-1,j,k}+V^t_{i+1,j,k}+ $\\ $V^t_{i,j+1,k}+V^t_{i+1,j,k}+V^t_{i,j+1,k})$\end{tabular} \\ \hline
5           & \begin{tabular}[c]{@{}c@{}}$C_1.V^t_{i,j-1,k}+C_2.V^t_{i-1,j,k}+C_3.V^t_{i,j,k-1}+$\\ $C_4.V^t_{i,j,k}+C_5.V^t_{i+1,j,k}+C_6.V^t_{i,j+1,k}$\end{tabular} \\ \hline
\end{tabular}
\caption{Stencil kernels.}
\label{tab:kernels}
\vspace{-0.35cm}
\end{table}

With all of these components in place, the proposed OpenMP runtime can to distribute tasks IPs across a cluster of FPGAs and map the dependency graph so that FPGA IPs communicate directly.

\section{An Stencil Multi-FPGA Pipeline}\label{sec:stencil}

Stencil computation is a method where a matrix (i.e. grid) is updated iteratively according to a fixed computation pattern \cite{roth:stencil:1997}. Stencil computations are used in this paper to show off the potential of the proposed OpenMP-based Multi-FPGA programming model. In this paper, stencil IPs are used to process multiple portions and  iterations of a grid  in parallel on different FPGAs. There are basically two types of parallelism that can be exploited when implementing stencil computation in hardware: \textit{cell-parallelism} and \textit{iteration-parallelism} \cite{waidyasooriya:fpgastencil:2019}.

As detailed below, these two types of parallelism leverage on a pipeline architecture to improve performance and are thus good candidates to take advantage of the Multi-FPGA programming model described herein. Five different types of stencil IPs have been implemented for evaluation. The IPs were adapted from \cite{waidyasooriya:fpgastencil:2019} and their computations are listed in Table \ref{tab:kernels} in the following order: (1) Laplace eq. 2-D, (2) Diffusion 2-D,  (3) Jacobi 9-pt. 2-D,   (4) Laplace eq. 3-D e (5) Diffusion 3-D. The formula in the \textit{computations} column is used to calculate an element $V^{t+1}_{i,j,k}$, where \textit{t} represents the iteration and the indices \textit{i}, \textit{j} and \textit{k} represent the axes of the grid. The $C_*$ values are constants passed to the IPs.

\begin{figure}[!t]
\centering
\begin{subfigure}[!t]{0.45\textwidth}
   \includegraphics[width=\columnwidth,height=4.25cm]{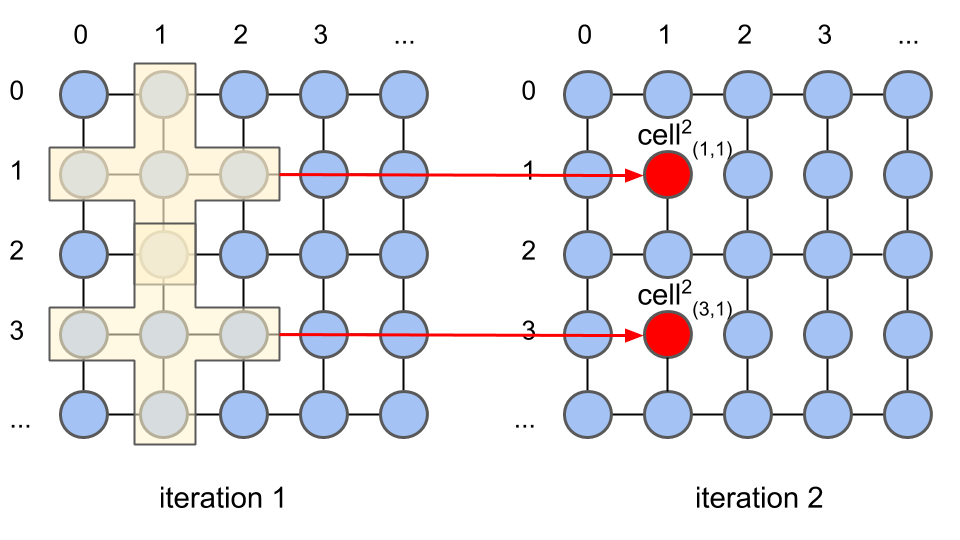}
   \caption{Cell-Parallel computation.}
   \label{fig:cell-parallel} 
\end{subfigure}

\begin{subfigure}[!h]{0.45\textwidth}
   \includegraphics[width=\columnwidth,height=4.25cm]{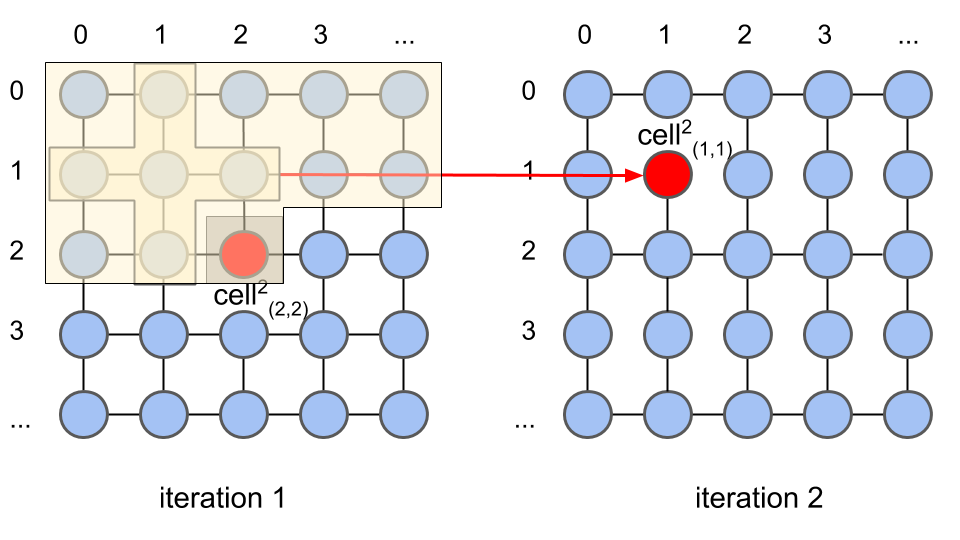}
   \caption{Iteration-Parallel computation.}
   \label{fig:iterations-parallel}
\end{subfigure}
\caption{Types of stencil parallelism. Adapted from \cite{waidyasooriya:fpgastencil:2019}.}
\end{figure}

%\label{subsec:cell}
\noindent{\bf Cell-Parallelism.}  Figure \ref{fig:cell-parallel} shows an example of  \textit{cell-parallelism} on a stencil computation, where  $cell^2_{(1,1)}$ at iteration 2 is computed using the data from its neighboring cells in the yellow area at iteration 1. This can be repeated for other cells at iteration 2,  like $cell^2_{(3,1)}$ which is  computed in parallel to  $cell^2_{(1,1)}$.

\begin{figure}[!h]
\centering
\begin{subfigure}[!h]{0.45\textwidth}
   \includegraphics[width=\columnwidth,height=4.5cm]{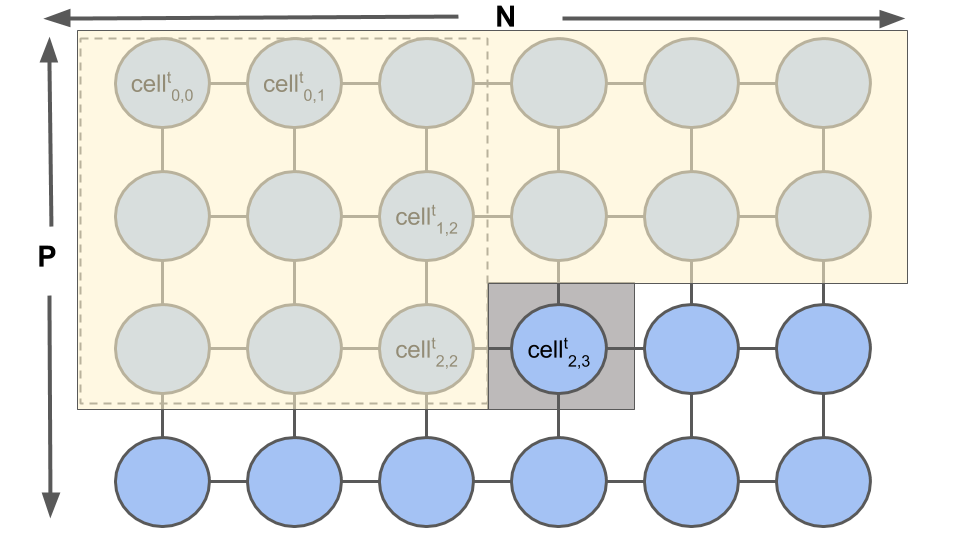}
   \caption{Grid Stencil to be computed.}
   \label{fig:shiftregistera} 
\end{subfigure}

\begin{subfigure}[!h]{0.45\textwidth}
   \includegraphics[width=\columnwidth,height=4.0cm]{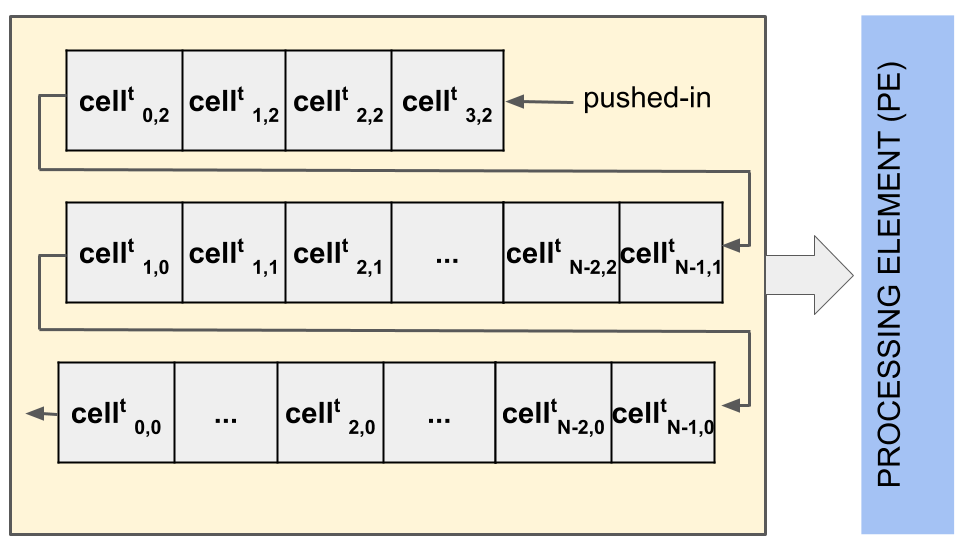}
   \caption{Shift-register and PE implementation.}
   \label{fig:shiftregisterb}
\end{subfigure}

\caption{IP implementation. Adapted from \cite{waidyasooriya:fpgastencil:2019}.}
\label{fig:shiftregister}
\vspace{-0.5cm}
\end{figure}

%\label{subsec:iterations}
\noindent{\bf Iteration-Parallelism}. This occurs when elements of different iterations are calculated in parallel. Figure \ref{fig:iterations-parallel} shows two consecutive iterations (1 and 2) where this happens. As shown in Figure \ref{fig:iterations-parallel} $cell^2_{(1,1)}$ at iteration 2 is computed using the data from its neighboring cells in the yellow area at iteration 1.  The method also computes other cells in parallel at iteration 2, like $cell^1_{(2,2)}$.

\subsection{IP Implementation}\label{subsec:ipstencil}

Figure \ref{fig:shiftregister} shows an overview of a typical stencil IP implementation using cell and iteration parallelism. Figure \ref{fig:shiftregistera} shows the grid to be computed, and Figure \ref{fig:shiftregisterb} the components that implement the stencil, namely: (a) a \textit{shift-register} that stores the grid data in processing order;  and (b) the \textit{processing element} (PE), which does the actual stencil computation. The cells in Figure \ref{fig:shiftregistera} are computed by the architecture in Figure \ref{fig:shiftregistera} from left-to-right and top-to-bottom one after the other. At each clock cycle, data in the shift-registers are shifted to the left in Figure \ref{fig:shiftregistera}, and a new cell value is pushed into the input of the first shift-register (i.e. $cell^t_{3,2}$). The computation starts after all neighboring data of a cell are available in the shift-register array. In the example of Figure \ref{fig:shiftregister},   $cell^{t+1}_{1,1}$ is computed while input  data is stored into $cell^t_{3,2}$. In the next clock cycle, the data at $cell^t_{0,0}$ at the output of the shift-register is discarded (shifted out), and the data of $cell^t_{4,2}$ is pushed into the input of the shift-register. Notice that the data at $cell^t_{0,0}$ is no longer required for any computation at this stage.

Each stencil IP has a \textit{shift-register} and eight processing elements and is thus capable of processing up to eight elements at a time until the end of an iteration. Each IP works with a 256-bit AXI4-Stream interface, as each cell in the matrix is a 32-bit float. 

The A-SWT switch in the architecture of  \rfig{cluster_inc} can be configured so that the IPs can be reused, thus expanding the system's capacity to deal with larger grids and iteration counts. By doing so,  the stencil pipeline can be scaled in both space and time. Such scaling is required to leverage the processing power of the  multiple FPGAs, and to enable the computation of large-size problems that could not be done by a single FPGA due to the lack of resources. Unfortunately, as discussed in the \rsec{experiments}, the size and number of IPs in an FPGA is constrained by the ability of the synthesis tool and designer to make efficient use of the FPGA resources, and this sometimes can become a bottleneck.

\section{Experimental Results}
\label{sec:experiments}

To evaluate the proposed system, three sets of experiments were performed using the stencil IPs described in Table \ref{tab:kernels}. The first set (\rsec{fgpascaling}) aimed at evaluating the scalability of the system with respect to the number of FPGAs. For the second set of experiments (\rsec{IPscaling}) the scalability concerning the number of IPs (i.e. number of iterations) was evaluated. Finally, the goal of the third set of experiments was to evaluate FPGA resource utilization. For all experiments, the board used was the Virtex-7 FPGA VC709 Connectivity Kit \cite{XilinxVi29:online} which contains a  Xilinx Virtex-7 XC7VX690T-2FFG1761C FPGA. Compilation of the HDL codes was done using Vivado 2018.3 \cite{VivadoDe67:online}.

\noindent{\bf Infra-structure issues.} The reader must keep in mind that the goal of these experiments was not to show raw performance numbers but to demonstrate the viability and scalability of the proposed programming model. Unfortunately, the infra-structure used in the experiments is not new. They have old Intel Xeon E5410 @2.33GHz CPUs, DDR2 667MHz memories, and archaic PCIe gen1 interfaces, which caused a considerable loss of performance since the FPGA boards use PCIe gen3. Moreover, as detailed in \rsec{resources}, the size of the original TRD kit made it very hard for Vivado to  synthesize  more IPs per FPGA, thus reducing the number of grid points inside the hardware, and the number of iterations. This harmed the final FPGA utilization and overall performance. However, even under these drawbacks the proposed approach still achieved linear speedups. Therefore, we are confident that after using more modern machines, FPGAs (e.g. U250) and design flow (e.g. Vitis), the resulting performance will be very competitive to that shown in the hand-designed solution of \cite{waidyasooriya:fpgastencil:2019}, which in some cases surpasses the performance of GTX 980 Ti and P100 GPUs.

\begin{table}[!t]
\begin{center}
\begin{tabular}{|c|c|c|c|}
\hline
Stencil Name     & Grid Size    & Iterations & \# IPs \\ \hline
Laplace 2D       & 4096x512    & 240        & 4             \\ \hline
Laplace 3D       & 512x64x64   & 240        & 2             \\ \hline
Difussion 2D     & 4096x512    & 240        & 1             \\ \hline
Difussion 3D     & 256x32x32   & 240        & 1             \\ \hline
Jacobi 9-pt. 2-D & 1024x128    & 240        & 1             \\ \hline
\end{tabular}
\caption{The setup of the stencil IPs.}
\label{tab:scal}
\end{center}
\vspace{-0.5cm}
\end{table}

\subsection{FPGA Scalability}
\label{sec:fgpascaling}

The FPGA's scalability experiments were executed with the settings shown in Table \ref{tab:scal}, and varying the number of FPGAs from 1 to 6. The \textit{Grid Size} column shows the dimensions of the initial grid for each kernel. The more computation a kernel does, the more difficult it was for Vivado to synthesize the design respecting the time constraints. For this reason, the dimensions of the grid at each kernel were adjusted to avoid negative slacks. The \textit{Iterations} column was set at 240 so that it was possible to execute with all  6 FPGAs. The \textit{\# IPs} column specifies the number of IPs at each FPGA. The number of IPs varies for the same reason as the dimensions of the initial grid. The  larger the kernel computation, the smaller the number of IPs Vivado could synthesize within the time constraints. On the other hand, as discussed in \rsec{resources} there is still plenty of hardware to be used before the FPGA runs out of resources, which reinforces the long term potential of the model proposed herein if using a more efficient FPGA design flow like (e.g. U250 and Vitis). 

\begin{figure}[!h]
    \centering
    \includegraphics[width=\columnwidth]{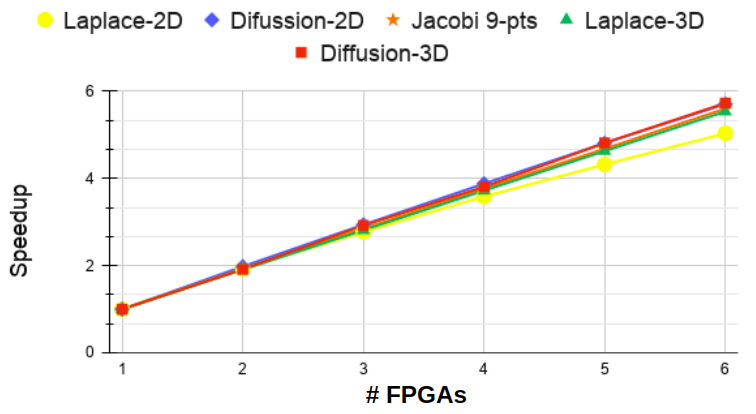}
    \caption{Speedup scaling with the number of FPGAs.}
    \label{fig:speedupxnfpgas}
\end{figure}

\begin{figure}[!h]
    \centering
    \includegraphics[width=\columnwidth]{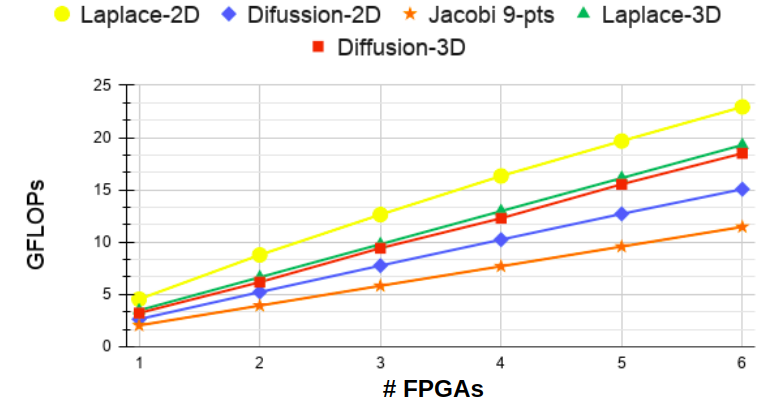}
   \caption{GFLOPS scaling with the number of FPGAs.}
    \label{fig:gflopsxnfpgas}
\end{figure}

The graph of Figure \ref{fig:speedupxnfpgas} shows the speedup concerning the execution on a single FPGA, achieved by the various stencil kernels as the number of FPGAs varies on the x-axis. The speedup grows almost linearly with the number of FPGAs for all five kernels. This result shows that it is possible to scale applications using Multi-FPGA architectures by using programming models like the one proposed in this paper to facilitate the design of such systems. 
The graph of Figure \ref{fig:gflopsxnfpgas} shows, on the y-axis, the number of floating-point operations (GFLOPs) for each kernel as the number of FPGA varies on the x-axis. The Laplace-2D kernel (yellow line) executes more GFLOPs than the other kernels. This is because, although the computation of this kernel is the simplest one, during synthesis, it was possible to insert more IPs (four) per FPGA, which allowed more iteration parallelism as discussed in \rsec{stencil}. Just below the Laplace-2D is the Laplace-3D (green line), with only 2 IPs per FPGA still managed to sustain a linear performance growth. For the remaining kernels, as they all have only one IP per FPGA, the number of GFLOPs is related to the number of operations executed and the grid's dimensions. Notice that Diffusion-3D (red line) and Diffusion-2D (blue line) perform less computation than the Jacobi 9-pts (orange line). However, they achieve better GFLOP numbers due to their higher grid dimension, enabling them to take advantage of increased iteration parallelism.


\begin{figure}[!h]
    \centering
    \includegraphics[width=\columnwidth]{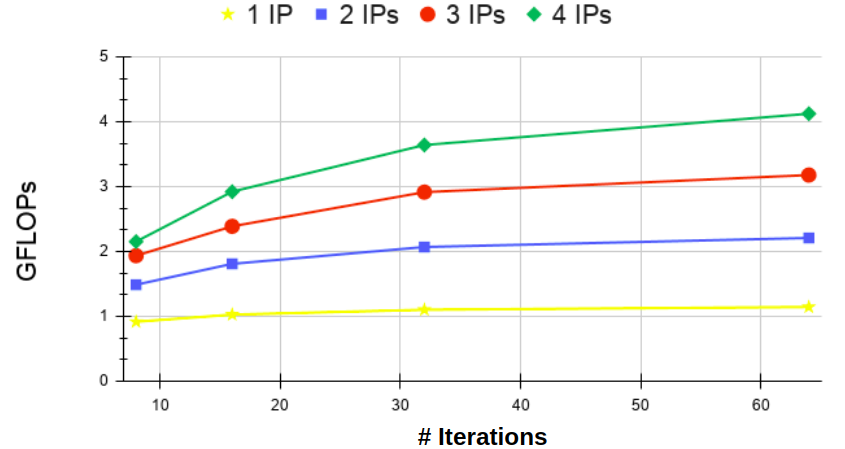}
    \caption{Laplace-2D scaling with the number of iterations.}
    \label{fig:gflopsxiterations}
\end{figure}
\begin{figure}[!h]
    \centering
    \includegraphics[width=\columnwidth]{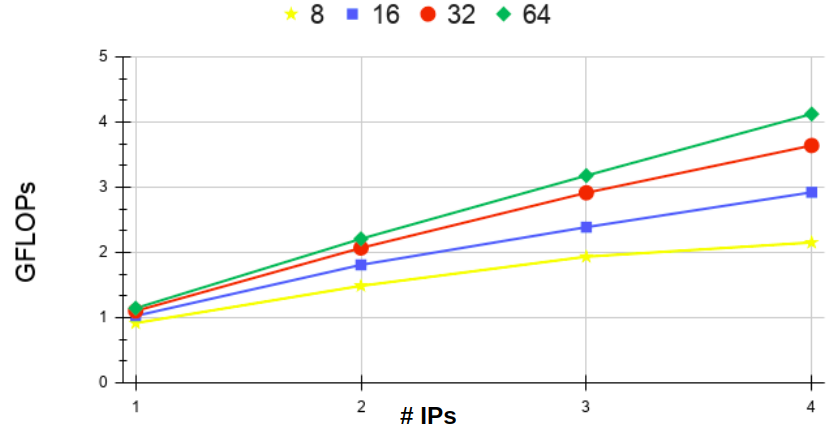}
    \caption{Laplace-2D scaling with the number of IPs.}
    \label{fig:gflopsxips}
\end{figure}

% paragraph explaining the experimental setup
% baseline table (single node): column 1 Stecils, 2 time, 
% scallability graph: x-axys; 1-8 FPGAs, Speed-up wrt baseline, vários stencils

\subsection{Iteration and IP Scalability}
\label{sec:IPscaling}

A second experiment was performed to evaluate the IPs' scalability concerning the number of iterations. The Laplace-2D kernel was used as an example, although similar results have been achieved for the other kernels. The graph in Figure \ref{fig:gflopsxiterations} shows, on the y-axis, the number of GFLOPs produced by the system, as the number of iterations varies on the x-axis. The yellow, blue, red, and green lines represent executions with 1, 2, 3, and 4 IPs, respectively. As shown, the execution with a single IP (yellow line) remains practically constant. 

On the other hand, the execution with 4 IPs shows an increase in performance until reaching a plateau. The executions with 2 and 3 IPs also show a gradual performance increase.  This experiment reveals that by increasing the number of IPs, it is possible to improve the system's scalability in terms of iterations.

The graph of Figure \ref{fig:gflopsxips} shows on the y-axis the number of GFLOPs for the Laplace-2D kernel as the number of IPs increase (x-axis). Each line in the graph is a different number of iterations. The graph reinforces the insight revealed in Figure \ref{fig:gflopsxiterations}: as more IPs are added to the system, the more significantly the increase in the number of iterations improve performance. This can be confirmed by looking at the distances between the lines in Figure \ref{fig:gflopsxips}, which grow larger as the number of IPs increase. This experiment also supports the case for Multi-FPGA architectures.

% Please add the following required packages to your document preamble:
% \usepackage{multirow}
\begin{table}[!t]
\begin{tabular}{|c|c|c|c|c|c|c|}
\hline
\multirow{2}{*}{Stencil} & \multicolumn{2}{c|}{Slice LUTs} & \multicolumn{2}{c|}{Block RAM} & \multicolumn{2}{c|}{DSP} \\ \cline{2-7} 
                         & \textit{\#}            & \%             & \textit{\#}           & \%             & \textit{\#}        & \%          \\ \hline
Laplace-2D               & 12138          & 7,5\%          & 8             & 0,7\%          & 16         & 0,4\%       \\ \hline
Diffusion-2D             & 25024          & 15,4\%         & 8             & 0,7\%          & 80         & 2,2\%       \\ \hline
Jacobi 9-pt              & 45733          & 28,3\%         & 8             & 0,7\%          & 144        & 4,0\%       \\ \hline
Laplace 3-D              & 21790          & 13,5\%         & 65            & 6,0\%          & 17         & 0,5\%       \\ \hline
Difussion-2D             & 27615          & 17,1\%         & 23            & 2,1\%          & 97         & 2,7\%       \\ \hline
\end{tabular}
\caption{IPs resource usage.}
\label{tab:ipuse}
%\vspace{-0.1cm}
\end{table}

\subsection{Resource Utilization}
\label{sec:resources}

\begin{figure}[!t]
    \centering
    \includegraphics[width=\columnwidth]{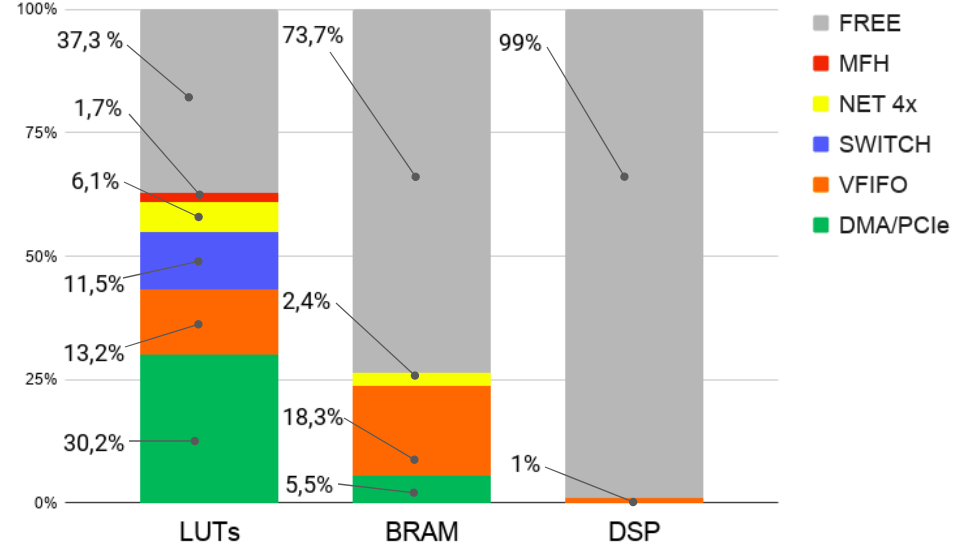}
    \caption{Resource usage distribution of the FPGA hardware.}
    \label{fig:ocupationchart}
    \vspace{-0.3cm}
\end{figure}

Regarding resource utilization, the graph in Figure \ref{fig:ocupationchart} shows the percentage of occupancy of the FPGA main components of the proposed infra-structure (not considering the IPs). Remarkably, the DMA/PCIe component occupies 30.2\% of the available LUTs. This large utilization is because the DMA/PCIe was designed to support a  board with four communication channels, although the proposed approach just requires one. Components MFH, SWITCH, VFIFO, and Network occupy, respectively, 1.7\%, 11.5\%, 13.2\% , and 6.1\% of the available LUTs. BRAMs are used by the DMA/PCIe (5.5\%), VFIFO (18.3\%), and NET (2.4\%). The most significant usage of BRAMs comes from VFIFO, which uses it to multiplex and demultiplex the four channels of the virtual FIFO. DSP is the least used component (1\%).

Table \ref{tab:ipuse} shows the quantity and  percentage of the FPGA components used by each IP from  the \textit{free} region (gray area) of \rfig{ocupationchart}. The percentage of the available LUTs effectively used by the stencil IPs varies from 7.5\% to 28.3\%, depending on the complexity of the kernel. As for BRAM, the utilization ranges from 0.7\% to 6.0\%. This is directly linked to the size of the \textit{shift-registers}, and is impacted by the size of the grid to be calculated. The number of DSP components used by the IPs varies from 0.4\% to 4.0\%, and is related to the number of multiplications performed by each kernel. The small utilization of the FPGA  resources by the kernels has been previously discussed in the beginning of \rsec{experiments}. Additional work will be done to address these shortcomings.

\section{Related Works} 
\label{sec:relwork}

This section summarizes the main works found in the literature with proposals for using OpenMP in FPGAs that uses LLVM. 

Choi et al \cite{choi:hls:2013} uses information  provided by pragmas to generate better parallel hardware. The compiler synthesizes one kernel IP per thread in the source program.

Sommer et al \cite{sommer:fpga:2017} uses Vivado HLS to generate hardware for the code regions annotated with OpenMP target directives. Their work fully supports \textit{omp target} directives  (including its \textit{map} clause). It is also the first work that leverages the LLVM libomptarget library to enable an FPGA synthesis flow. 

%\subsection{Ceissler* (2018)} \label{subsec:ceissler}

Ceissler et al \cite{ceissler:hardcloud:2018} describe  HardCloud, an OpenMP platform that integrates pre-designed FPGA IPs  to an  OpenMP program. The authors propose three more clauses to the OpenMP 4.X standard: \textit{use}, \textit{check} and \textit{module} that allow the access  of IP's inputs/outputs directly from  OpenMP code.

Knaust et al \cite{knaust:openclsdk:2019} use Clang \cite{clang} to outline \textit{omp target} regions at the level of the LLVM IR, and feed them into Intel’s OpenCL HLS tool-chain to generate a hardware kernel for the FPGA. Their approach uses Intel’s OpenCL API to allow the communication between host and FPGA. 

To the best of our knowledge, and contrary to the previous works, which focused mostly on the synthesis and single FPGA architectures, this paper is the first to enable OpenMP task parallelism to integrate IPs into a Multi-FPGA architecture.

\section{Conclusions}
\label{sec:conclusions}

This paper proposes to extend the OpenMP task-based computation offloading programming model to enable several FPGAs to work together as a single Multi-FPGA architecture. Experimental results for a set of OpenMP stencil-based applications running on a Multi-FPGA platform consisting of 6 Xilinx VC709 FPGA boards interconnected through fiber-optic links have shown close to linear speedups as the number of FPGAs and IP-cores per FPGA increase.

\newpage
\bibliographystyle{IEEEtran}
\bibliography{bare_conf}

% Generated by IEEEtran.bst, version: 1.14 (2015/08/26)
\begin{thebibliography}{10}
\providecommand{\url}[1]{#1}
\csname url@samestyle\endcsname
\providecommand{\newblock}{\relax}
\providecommand{\bibinfo}[2]{#2}
\providecommand{\BIBentrySTDinterwordspacing}{\spaceskip=0pt\relax}
\providecommand{\BIBentryALTinterwordstretchfactor}{4}
\providecommand{\BIBentryALTinterwordspacing}{\spaceskip=\fontdimen2\font plus
\BIBentryALTinterwordstretchfactor\fontdimen3\font minus
  \fontdimen4\font\relax}
\providecommand{\BIBforeignlanguage}[2]{{%
\expandafter\ifx\csname l@#1\endcsname\relax
\typeout{** WARNING: IEEEtran.bst: No hyphenation pattern has been}%
\typeout{** loaded for the language `#1'. Using the pattern for}%
\typeout{** the default language instead.}%
\else
\language=\csname l@#1\endcsname
\fi
#2}}
\providecommand{\BIBdecl}{\relax}
\BIBdecl

\bibitem{oneal:surveypower:2018}
K.~{O'Neal} and P.~{Brisk}, ``Predictive modeling for cpu, gpu, and fpga
  performance and power consumption: A survey,'' in \emph{2018 IEEE Computer
  Society Annual Symposium on VLSI (ISVLSI)}, July 2018, pp. 763--768.

\bibitem{guo:gpu:2020}
Z.~{Guo}, T.~W. {Huang}, and Y.~{Lin}, ``Gpu-accelerated static timing
  analysis,'' in \emph{2020 IEEE/ACM International Conference On Computer Aided
  Design (ICCAD)}, 2020, pp. 1--9.

\bibitem{ghazawi:promise:2008}
T.~{El-Ghazawi}, E.~{El-Araby}, M.~{Huang}, K.~{Gaj}, V.~{Kindratenko}, and
  D.~{Buell}, ``The promise of high-performance reconfigurable computing,''
  \emph{Computer}, vol.~41, no.~2, pp. 69--76, Feb 2008.

\bibitem{lee:openaccfpga:2016}
S.~{Lee}, J.~{Kim}, and J.~S. {Vetter}, ``Openacc to fpga: A framework for
  directive-based high-performance reconfigurable computing,'' in \emph{2016
  IEEE International Parallel and Distributed Processing Symposium (IPDPS)},
  May 2016, pp. 544--554.

\bibitem{strickland:fpgahpc:2018}
M.~{Strickland}, ``Fpga accelerated hpc and data analytics,'' in \emph{2018
  International Conference on Field-Programmable Technology (FPT)}, Dec 2018,
  pp. 21--21.

\bibitem{reichenbach:hetfpga:2019}
\BIBentryALTinterwordspacing
M.~Reichenbach, P.~Holzinger, K.~H{\"a}ublein, T.~Lieske, P.~Blinzer, and
  D.~Fey, ``Heterogeneous computing utilizing fpgas,'' \emph{Journal of Signal
  Processing Systems}, vol.~91, no.~7, pp. 745--757, Jul 2019. [Online].
  Available: \url{https://doi.org/10.1007/s11265-018-1382-7}
\BIBentrySTDinterwordspacing

\bibitem{caulfield:azure:2016}
A.~M. {Caulfield}, E.~S. {Chung}, A.~{Putnam}, H.~{Angepat}, J.~{Fowers},
  M.~{Haselman}, S.~{Heil}, M.~{Humphrey}, P.~{Kaur}, J.~{Kim}, D.~{Lo},
  T.~{Massengill}, K.~{Ovtcharov}, M.~{Papamichael}, L.~{Woods}, S.~{Lanka},
  D.~{Chiou}, and D.~{Burger}, ``A cloud-scale acceleration architecture,'' in
  \emph{2016 49th Annual IEEE/ACM International Symposium on Microarchitecture
  (MICRO)}, Oct 2016, pp. 1--13.

\bibitem{awsf1}
\BIBentryALTinterwordspacing
``{Amazon EC2 F1 Instances},''
  \url{https://aws.amazon.com/ec2/instance-types/f1}, Nov 2019, [Online;
  accessed 25. Nov. 2019]. [Online]. Available:
  \url{https://aws.amazon.com/ec2/instance-types/f1}
\BIBentrySTDinterwordspacing

\bibitem{cong:fxg:2018}
\BIBentryALTinterwordspacing
J.~Cong, Z.~Fang, M.~Lo, H.~Wang, J.~Xu, and S.~Zhang, ``Understanding
  performance differences of fpgas and gpus: (abtract only),'' in
  \emph{Proceedings of the 2018 ACM/SIGDA International Symposium on
  Field-Programmable Gate Arrays}, ser. FPGA '18.\hskip 1em plus 0.5em minus
  0.4em\relax New York, NY, USA: ACM, 2018, pp. 288--288. [Online]. Available:
  \url{http://doi.acm.org/10.1145/3174243.3174970}
\BIBentrySTDinterwordspacing

\bibitem{jiang:dnnfpgas:2019}
\BIBentryALTinterwordspacing
W.~Jiang, E.~H.-M. Sha, X.~Zhang, L.~Yang, Q.~Zhuge, Y.~Shi, and J.~Hu,
  ``Achieving super-linear speedup across multi-fpga for real-time dnn
  inference,'' \emph{ACM Trans. Embed. Comput. Syst.}, vol.~18, no.~5s, pp.
  67:1--67:23, Oct. 2019. [Online]. Available:
  \url{http://doi.acm.org/10.1145/3358192}
\BIBentrySTDinterwordspacing

\bibitem{farooq:routing:2018}
U.~{Farooq}, I.~{Baig}, and B.~A. {Alzahrani}, ``An efficient inter-fpga
  routing exploration environment for multi-fpga systems,'' \emph{IEEE Access},
  vol.~6, pp. 56\,301--56\,310, 2018.

\bibitem{azeem:fpgasdebug:2016}
M.~M. {Azeem}, R.~{Chotin-Avot}, U.~{Farooq}, M.~{Ravoson}, and H.~{Mehrez},
  ``Multiple fpgas based prototyping and debugging with complete design flow,''
  in \emph{2016 11th International Design Test Symposium (IDT)}, Dec 2016, pp.
  171--176.

\bibitem{waidyasooriya:fpgastencil:2019}
H.~M. {Waidyasooriya} and M.~{Hariyama}, ``Multi-fpga accelerator architecture
  for stencil computation exploiting spacial and temporal scalability,''
  \emph{IEEE Access}, vol.~7, pp. 53\,188--53\,201, 2019.

\bibitem{kunzman:proghet:2011}
D.~M. {Kunzman} and L.~V. {Kale}, ``Programming heterogeneous systems,'' in
  \emph{2011 IEEE International Symposium on Parallel and Distributed
  Processing Workshops and Phd Forum}, May 2011, pp. 2061--2064.

\bibitem{pu:prgdsl:2017}
\BIBentryALTinterwordspacing
J.~Pu, S.~Bell, X.~Yang, J.~Setter, S.~Richardson, J.~Ragan-Kelley, and
  M.~Horowitz, ``Programming heterogeneous systems from an image processing
  dsl,'' \emph{ACM Trans. Archit. Code Optim.}, vol.~14, no.~3, pp.
  26:1--26:25, Aug. 2017. [Online]. Available:
  \url{http://doi.acm.org/10.1145/3107953}
\BIBentrySTDinterwordspacing

\bibitem{openmp45}
\BIBentryALTinterwordspacing
``{OpenMP 4.5 Specifications},''
  \url{http://www.openmp.org/mp-documents/openmp-4.5.pdf}, {A}ccessed on
  Oct~13, 2019. [Online]. Available:
  \url{http://www.openmp.org/mp-documents/openmp-4.5.pdf}
\BIBentrySTDinterwordspacing

\bibitem{openmp2008book}
B.~Chapman, G.~Jost, and R.~Van Der~Pas, \emph{{Using OpenMP: Portable Shared
  Memory Parallel Programming}}.\hskip 1em plus 0.5em minus 0.4em\relax MIT
  press, 2008, vol.~10.

\bibitem{openmp30}
\BIBentryALTinterwordspacing
``{OpenMP 3.0 Specifications},''
  \url{https://www.openmp.org/wp-content/uploads/spec30.pdf}, {A}ccessed on
  Oct~13, 2019. [Online]. Available:
  \url{https://www.openmp.org/wp-content/uploads/spec30.pdf}
\BIBentrySTDinterwordspacing

\bibitem{budruk:PES:2003}
R.~Budruk, D.~Anderson, and E.~Solari, \emph{PCI Express System
  Architecture}.\hskip 1em plus 0.5em minus 0.4em\relax Pearson Education,
  2003.

\bibitem{lattner:LCF:2004}
\BIBentryALTinterwordspacing
C.~Lattner and V.~Adve, ``Llvm: A compilation framework for lifelong program
  analysis \& transformation,'' in \emph{Proceedings of the International
  Symposium on Code Generation and Optimization: Feedback-directed and Runtime
  Optimization}, ser. CGO '04.\hskip 1em plus 0.5em minus 0.4em\relax
  Washington, DC, USA: IEEE Computer Society, 2004, pp. 75--. [Online].
  Available: \url{http://dl.acm.org/citation.cfm?id=977395.977673}
\BIBentrySTDinterwordspacing

\bibitem{bertolli:IGS:2015}
\BIBentryALTinterwordspacing
C.~Bertolli, S.~F. Antao, G.-T. Bercea, A.~C. Jacob, A.~E. Eichenberger,
  T.~Chen, Z.~Sura, H.~Sung, G.~Rokos, D.~Appelhans, and K.~O'Brien,
  ``Integrating gpu support for openmp offloading directives into clang,'' in
  \emph{Proceedings of the Second Workshop on the LLVM Compiler Infrastructure
  in HPC}, ser. LLVM '15.\hskip 1em plus 0.5em minus 0.4em\relax New York, NY,
  USA: ACM, 2015, pp. 5:1--5:11. [Online]. Available:
  \url{http://doi.acm.org/10.1145/2833157.2833161}
\BIBentrySTDinterwordspacing

\bibitem{AXI4Stre90:online}
\BIBentryALTinterwordspacing
``Axi4-stream interconnect v1.1 logicore ip product guide (pg035).'' [Online].
  Available:
  \url{https://www.xilinx.com/support/documentation/ip_documentation/axis_interconnect/v1_1/pg035_axis_interconnect.pdf}
\BIBentrySTDinterwordspacing

\bibitem{roth:stencil:1997}
\BIBentryALTinterwordspacing
G.~Roth, J.~Mellor-Crummey, K.~Kennedy, and R.~G. Brickner, ``Compiling
  stencils in high performance fortran,'' in \emph{Proceedings of the 1997
  ACM/IEEE Conference on Supercomputing}, ser. SC '97.\hskip 1em plus 0.5em
  minus 0.4em\relax New York, NY, USA: Association for Computing Machinery,
  1997, p. 1–20. [Online]. Available:
  \url{https://doi.org/10.1145/509593.509605}
\BIBentrySTDinterwordspacing

\bibitem{XilinxVi29:online}
``Xilinx virtex-7 fpga vc709 connectivity kit,''
  \url{https://www.xilinx.com/products/boards-and-kits/dk-v7-vc709-g.html},
  (Accessed on 11/11/2020).

\bibitem{VivadoDe67:online}
``Vivado design suite user guide: Release notes, installation, and licensing
  (ug973),'' (Accessed on 11/16/2020).

\bibitem{choi:hls:2013}
J.~{Choi}, S.~{Brown}, and J.~{Anderson}, ``From software threads to parallel
  hardware in high-level synthesis for fpgas,'' in \emph{2013 International
  Conference on Field-Programmable Technology (FPT)}, Dec 2013, pp. 270--277.

\bibitem{sommer:fpga:2017}
L.~{Sommer}, J.~{Korinth}, and A.~{Koch}, ``Openmp device offloading to fpga
  accelerators,'' in \emph{2017 IEEE 28th International Conference on
  Application-specific Systems, Architectures and Processors (ASAP)}, July
  2017, pp. 201--205.

\bibitem{ceissler:hardcloud:2018}
C.~{Ceissler}, R.~{Nepomuceno}, M.~{Pereira}, and G.~{Araujo}, ``Automatic
  offloading of cluster accelerators,'' in \emph{2018 IEEE 26th Annual
  International Symposium on Field-Programmable Custom Computing Machines
  (FCCM)}, April 2018, pp. 224--224.

\bibitem{knaust:openclsdk:2019}
M.~{Knaust}, F.~{Mayer}, and T.~{Steinke}, ``Openmp to fpga offloading
  prototype using opencl sdk,'' in \emph{2019 IEEE International Parallel and
  Distributed Processing Symposium Workshops (IPDPSW)}, May 2019, pp. 387--390.

\bibitem{clang}
\BIBentryALTinterwordspacing
{LLVM Project}. (2007) {Clang: a C language family frontend for LLVM}.
  [Online]. Available: \url{https://clang.llvm.org/}
\BIBentrySTDinterwordspacing

\end{thebibliography}
% argument is your BibTeX string definitions and bibliography database(s)
%\bibliography{IEEEabrv,../bib/paper}
%
% <OR> manually copy in the resultant .bbl file
% set second argument of \begin to the number of references
% (used to reserve space for the reference number labels box)
%\begin{thebibliography}{1}

%\bibitem{IEEEhowto:kopka}
%H.~Kopka and P.~W. Daly, \emph{A Guide to \LaTeX}, 3rd~ed.\hskip 1em plus
%  0.5em minus 0.4em\relax Harlow, England: Addison-Wesley, 1999.

%\end{thebibliography}

% that's all folks
\end{document}